**This is the accepted manuscript (postprint) of the following article:**
N. Esmati, T. Khodaei, E. Salahinejad, E. Sharifi, *Fluoride doping into $SiO_2$-MgO-CaO bioactive glass nanoparticles: bioactivity, biodegradation and biocompatibility assessments*, Ceramics International, 44 (2018) 17506-17513.
https://doi.org/10.1016/j.ceramint.2018.06.222

# Fluoride doping into $SiO_2$-MgO-CaO bioactive glass nanoparticles: bioactivity, biodegradation and biocompatibility assessments

N. Esmati [a], T. Khodaei [a], E. Salahinejad [a,*], E. Sharifi [b]

[a] Faculty of Materials Science and Engineering, K. N. Toosi University of Technology, Tehran, Iran

[b] Cellular and Molecular Research Center, Basic Health Sciences Institute, Shahrekord University of Medical Sciences, Shahrekord, Iran

## Abstract

In this research, for the first time, the structure, bioactivity, biodegradation and biocompatibility of $SiO_2$-MgO-CaO glasses doped with different levels of fluoride were studied. The glassy powder samples were synthesized by a coprecipitation method followed by calcination at 500 °C, where amorphicity and fluoride incorporation were verified by X-ray diffraction and Raman spectroscopy, respectively. The *in vitro* biomineralization and biodegradation of the samples were also investigated by electron microscopy, Raman spectroscopy and inductively coupled plasma optical emission spectrometry. These assessments revealed that there is an optimum level of fluoride doping to meet the highest bioactivity. Remarkably, the same level of incorporation presented the foremost biocompatibility with respect to osteoblast-like MG-63 human cells, as realized by the MTT assay and cell attachment studies.

* Corresponding Author: Email Address: <salahinejad@kntu.ac.ir>

## 1. Introduction

Biomaterials with the ability to substitute or regenerate damaged tissues have attracted great attention in recent decades. In this regard, $SiO_2$-CaO-$Na_2O$-$P_2O_5$ glasses were originally introduced by Hench and Poldak [1] as the third generation of biomaterials, with bioactive and bioresorbable features. In the further development of such biomaterials, the incorporation of magnesium in glasses was regarded because this ion is beneficial for cell interactions [2-5]. One of the typical groups of magnesium-containing glasses is related to the $SiO_2$-MgO-CaO system, which is mainly known because of improved mechanical properties and osteoinductivity [5].

As well as the modification of basic composition, another approach to improving the biological performance of bioactive glasses is doping with proper ions. Fluoride is one of these dopants which is known to enhance the chemical stability and bioactivity of hard tissues in acidic environments, especially for dental applications [6-8]. In accordance with the literature, this dopant has been frequently used in some crystalline ceramics and glasses in biomedical areas. Typically, it has been reported that the incorporation of fluoride into crystalline diopside ($CaMgSi_2O_6$) improves its sinterability and apatite-forming ability [9, 10]. Another research conducted by Fahami et al. [11] reveals the alteration of the osseointegration and bone regeneration of hydroxyapatite after fluoride introduction. Also, the addition of fluoride to $Na_2O$-CaO-$SiO_2$-$P_2O_5$ glass ceramics has resulted in improved bioactivity and biocompatibility [12]. In a study reported by Brauer et al. [6], it has been also concluded that fluoride-containing $Na_2O$-CaO-$SiO_2$-$P_2O_5$ glasses can stimulate the deposition of fluorapatite and enhance bioactivity, whereas the higher amounts of fluoride results in the formation of fluorite and the reduction of biomineralization.

Overall, from the literature, it is concluded that there is an optimal value for fluoride doping to meet the best biological behaviors. In this regard, to the best of our knowledge, no study has been reported on fluoride doping in $SiO_2$-MgO-CaO bioactive glasses. Thus, in this research, a bioactive glass with the ionic composition of $2SiO_2$-MgO-CaO was doped with various amounts of fluoride. Then, the resultant structure, biodegradation, bioactivity and biocompatibility of the samples were compared *in vitro*.

## 2. Materials and methods

$2SiO_2$-MgO-CaO powders were synthesized by a coprecipitation method using chloride precursors, as followed. Silicon tetrachloride ($SiCl_4$, Merck, Germany, >99%), calcium chloride ($CaCl_2$, Merck, Germany, >98%) and magnesium chloride ($MgCl_2$, Merck, Germany, >98%) were used as the silicon, calcium and magnesium precursors, respectively. Also, magnesium fluoride ($MgF_2$, Alfa Aesar, >98%) was employed to incorporate fluoride as the dopant into the hosted substance. In addition, dry ethanol ($C_2H_5OH$, Merck, Germany, 99.9%) and ammonia solution ($NH_4OH$, Merck, Germany, 25%) were used as the solvent and precipitant, respectively.

According to Table 1, the given amounts of magnesium chloride and calcium chloride were added to 100 ml ethanol and stirred on a magnetic stirrer. The solution was then placed in a water-ice bath, followed by the addition of silicon tetrachloride. Afterwards, magnesium fluoride was added to the above solution, where the total molar ratios of Si:Mg:Ca were kept at 2:1:1 in all the samples. Finally, ammonia solution was added dropwise until a pH value of 10, whereas a white deposit was obtained due to a coprecipitation reaction. The obtained deposits were dried at 100 °C, ground in a manual aggregate mortar, and finally calcined at 500 °C to evacuate volatile coproducts. The molar percent of fluoride in the calcined samples was also listed in Table 1, as calculated by the amounts of the precursors used. The calcined powders were structurally characterized by X-ray diffraction (XRD) and Raman spectroscopy for crystallinity and bonding assessments, respectively.

In order to evaluate the bioactivity of the samples, 0.1 g of each powder sample was immersed in 30 ml of the simulated body fluid (SBF) for seven days at 37 °C. The pH value of the SBF exposed to the samples was measured using a digital pH meter every day. After 7 days of incubation, the powders were separated from the SBF, rinsed thoroughly with distilled water, dried at 100 °C, and characterized by scanning electron microcopy (SEM) equipped with energy-dispersive X-ray spectroscopy (EDS) and Raman spectroscopy. The exposed SBFs were also analyzed by inductively coupled plasma optical emission spectrometry (ICP).

For the *in vitro* biocompatibility assessment of the synthesized samples, the powders were first pressed into 5 mm diameter discs. The sterilization process of the discs was followed by immersion in an ethanol solution (70%), washing with phosphate-buffered saline, and UV exposure. $2*10^4$ osteoblast-like MG-63 cells were cultured on the samples in the humidified medium containing $CO_2$ for specific durations. The assessment of cytocompatibility was conducted by the MTT assay with three repetitions. Also, in order to investigate the cell attachment on the surfaces, after the cell fixation based on the protocol of Ref. [13], the samples were observed by SEM.

## 3. Results and discussion

The XRD pattern of the synthesized powder samples is shown in Fig. 1. According to this analysis, the samples calcined at 500 °C are completely amorphous, without any diffractions of crystalline phases. It is noteworthy that the calcination of these glassy powders at 700 °C has resulted in the appearance of the XRD peaks of diopside ($CaMgSi_2O_6$) [14-16].

Fig. 2 presents the Raman spectra of the powder samples calcined at 500 °C in the wavenumbers of 200-1600 $cm^{-1}$. The peak at 363 $cm^{-1}$ in all of the samples is assigned to the Ca-O stretching mode [17]. The peaks of 430, 490 and 620 $cm^{-1}$, which are also present for all of the samples, is related to the Si–O–Si linkage [18, 19]. The reduction in the intensity of these latter peaks as a result of fluoride-doping is an evidence for the incorporation of this species, which is also compatible with Ref. [18]. The appearance of a weak peak at 798 $cm^{-1}$, which is more typical for G2, is also attributed to the formation of the Si-F bond [20]. The introduction of fluoride into crystalline diopside synthesized by the same coprecipitation process, but calcined at temperatures higher than 500 °C, has been previously proved by Fourier-transform infrared spectroscopy [9, 10]. Considering that the redius of $F^-$ is close to that of $O^{2-}$, each $O^{2-}$ in the slicitae chane is partialy replaced with two $F^-$ to keep the electrostatic charage balance. This means that each bridging oxygen (Si-O-Si) is substituted by two non-bridging fluorides (Si-F + F-Si) [10, 21].

The SEM micrographs of the selected powder samples after calcination are demonstrated in Fig. 3. It can be seen that the particles mostly are agglomerated, equiaxed and irregular in shape with a mean size of almost 70 nm. Fig. 4 also presents the micrographs of the selected powders after 7 days of immersion in the SBF. As a result of immersion in the SBF, new precipitates observed in a brighter contrast have been formed on the powder surfaces appearing in a darker contrast. The number and size of the precipitates increase from G0 to G1, but decrease from G1 to G2. The presence of phosphorus in the EDS spectra taken of the precipitates suggests that they are composed of apatite (calcium phosphate). Thus, it is concluded that G1 presents a better bioactivity (apatite-formation ability) compared to G0 and G2, based on the SEM-EDS studies.

To further characterize the apatite precipitates formed after incubation in the SBF, Raman spectroscopy was used on the immersed powders (Fig. 5). In comparison to Fig. 2, the intensity of the 363 $cm^{-1}$ peak (the Ca-O bond) was enhanced after immersion, especially for G0.5, G1, and G1.5, suggesting the precipitation of a Ca-containing species (apatite). The peak at 709 $cm^{-1}$ is also assigned to the formation of Ca-$CO_3$ [22]. The peak at 1081 $cm^{-1}$ for the F-containing samples is attributed to the incorporation of fluoride into apatite [23]. Thus, it is concluded that the precipitates formed due to immersion in the SBF are apatite. In addition, apatite formed on G0.5, G1, and G1.5 is incorporated with fluoride and is higher than G0 and G2 in amount, which is compatible with the SEM evaluations. Indeed, the introduction of fluoride into apatite increases its chemical stability against dissolution in the SBF [22], which explains the higher bioactivity of G0.5, G1, and G1.5.

The ICP analysis was also done on the SBF in contact to the samples after 7 days of incubation (Fig. 6). The ionic concentration of phosphorous in the SBF is the most key factor for comparing the bioactivity of the samples. The reduction in the phosphorous level of the SBF after immersion suggests its adsorption on the powder surfaces and the precipitation of a calcium phosphate layer, as explored in the SEM and Raman evaluations. This is an essential evidence for *in vitro* and *in vivo* bioactivity [24]. In this regard, the SBF associated to G1 and G1.5 shows the lowest phosphorous contents, indicating the highest apatite-formation ability. In contrast, G2 yields the highest phosphorous content in the SBF and thereby the lowest bioactivity. It is noticeable that the bioactivity ranking realized by ICP is completely compatible with the SEM and Raman results. The dissolution of calcium and magnesium from the samples toward the SBF is also a key step in the precipitation of apatite, via leaving a silica-gel layer on the surface. Then, calcium and phosphorous are adsorbed on this gel layer and gradually form an apatite layer. The decline in the magnesium and silicon contents in the exposing SBF, compared to the fresh SBF, is also a evidence for biodegradation. Concerning fluoride released from the samples toward the SBF, there are two competitive consumers depending on the fluoride level. Fluoride at low contents, on the one hand, can be incorporated into apatite and enhance its stability [22], explaining the increase in bioactivity from G0 to G1. On the other hand, fluoride released at high levels can contribute to the formation of fluorite ($CaF_2$) as a competitive user of calcium [10], decreasing apatite-formation ability as observed for G2.

The pH variation of the SBF in contact with the powders is also exhibited in Fig. 7. For all of the samples, pH is enhanced within the first days of immersion and then presents a decrease to the 7$^{th}$ day of incubation. The primary increase is due to the fast release of $Mg^{2+}$ and $Ca^{2+}$ from the samples into the SBF, which causes the adsorption of $H^+$ or $H_3O^+$ from the SBF on the surface. During these ion-exchange reactions, the acidic factor in the SBF decreases and pH is therefore enhanced. Nevertheless, the subsequent decrease of pH is a result of the formation of apatite layers on the powder surfaces since hydroxyapatite ($Ca_5(PO_4)_3(OH)$) consumes $OH^-$ for formation. This decrease is more pronounced for G1 and G1.5 which represent the most apatite-formation ability based on the above characterizations. Regarding the effect of fluoride, it can be seen over the entire range of incubation that the fluoride-containing samples yield lower pH values in comparison to G0. Because ion exchanges between $F^-$ from the samples and $OH^-$ from the SBF result in the deficiency of the basic factor in the SBF, lowering pH.

Fig. 8 shows the MTT results of the MG-63 cell cultures on the pressed powders. As it can be seen, at 24 h, the optical densities of viable cells (OD) for the doped samples are comparable with that of G0 which is regarded as the control because the substantial biocompatibility of diopside, i.e. the crystalline form of G0, has been previously reported [25, 26]. It suggests the cytocompatibility of all the samples. The increase in the OD values with increasing the culture time from 24 h to 72 h points out the cell proliferation on the surfaces. As can be seen, the cell viability (relative OD with respect to the control) at 72 h of culture is meaningfully improved by increasing the content of fluoride from 0 to 1 %, reaches a maximum for G1 and G1.5, and finally is decreased for G2. Remarkably, G1 and G1.5 simultaneously exhibit the highest bioactivity and biocompatibility; that is, 1 and 1.5 % are the optimal incorporation levels of fluoride into the $2SiO_2$-MgO-CaO glass. The evaluation of the cell attachment on the surfaces also confirms the MTT results (Fig. 9). Typically, the cells cultured on the G0 surface are separate from each other, but they display relatively broad lamellipodia with a number of protrusive filopodia. The cell spreading on G1 is so mature that the cells are connected to each other and cover a considerable area of the surface. However, the evolution of the cell attachment on G2 is comparatively undeveloped.

Based on the literature, the influence of fluoride on biocompatibility is determined by the level of fluoride released into the medium and the type of the cultured cells. The released concentration depends on the type of the hosted bioceramic, the amount of fluoride doped into the ceramic, and the culture duration. At concentrations below a critical value of release, fluoride encourages cell activity; however, a deteriorous effect is obtained at higher values [27, 28]. Thus, it is concluded that in this work, the activity of MG-63 cells to 3 days of culture benefits from the addition of fluoride to 1%; however, a toxic effect is achieved beyond 1.5% of incorporation, the case occurring for G2. Apart from the direct effect of fluoride on the cell activity, the progress of biocompatible apatite precipitation on the surfaces (Figs. 4 to 6) and pH buffering (Fig. 7), by increasing the content of fluoride to 1 %, can be effective on the improved biocompatibility.

## 4. Conclusions

Fluoride up to 2 mol.% was successfully doped into $2SiO_2$-MgO-CaO glass nanopowders by a coprecipitation process using inorganic precursors, followed by calcination at 500 °C. It was realized that by increasing the fluoride level to 1 %, the bioactivity was *in vitro* improved. This characteristic reached a maximum at 1 and 1.5 % of incorporation, but then decreased for the 2% sample. Multiple ion-exchange reactions were explored to control the related biomineralization behavior. The response of osteoblast-like MG-63 cells to 72 days of culture suggested a biocompatible behavior for the prepared samples, where the highest cell proliferation was found to for 1-1.5 % of addition. It is eventually concluded that the fluoride incorporation of 1-1.5 % was the optimal level to meet the highest bioactivity and biocompatibility *in vitro*.

## Figures

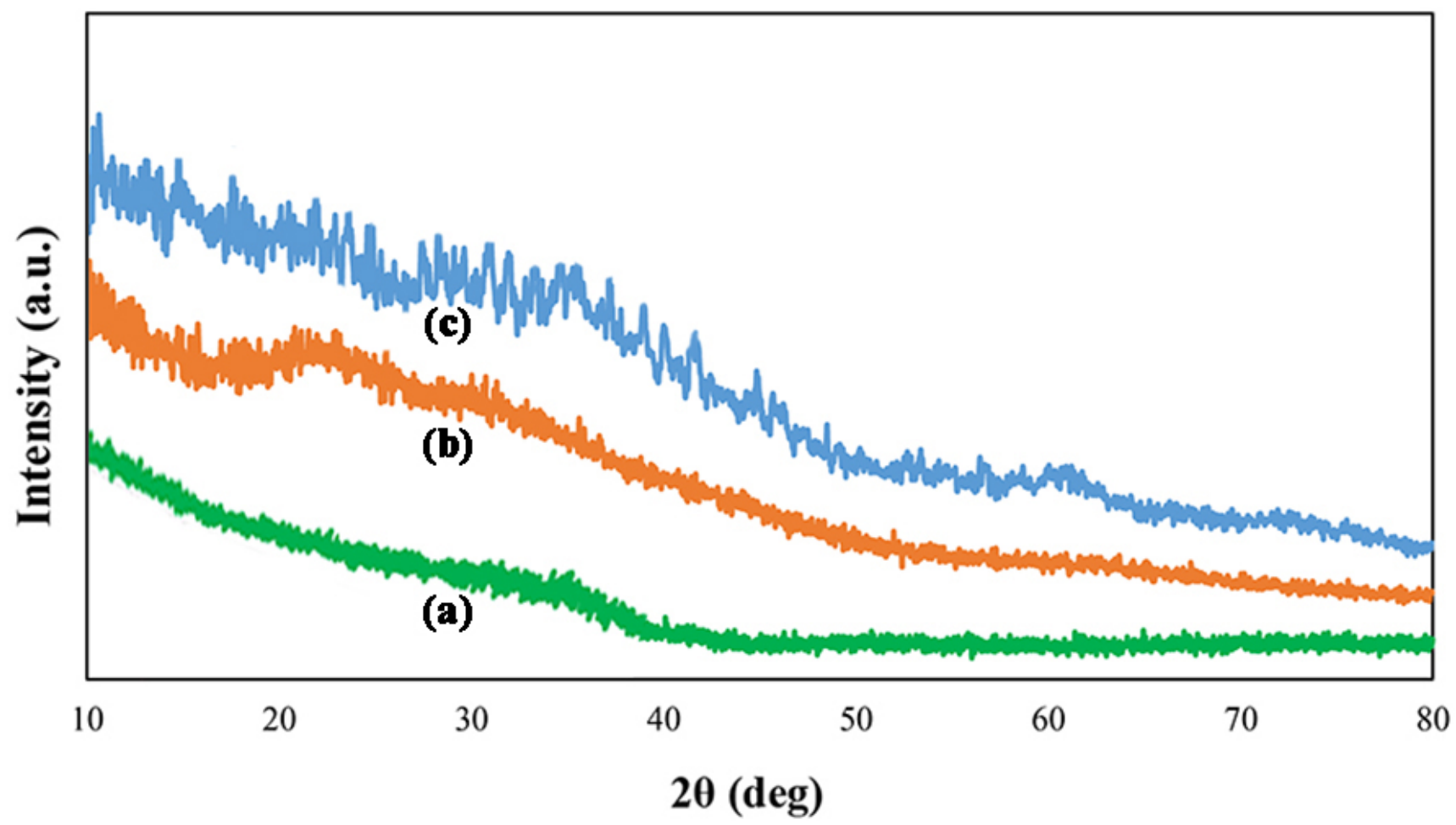

Fig. 1. XRD patterns of the calcined samples: G0 (a), G1 (b) and G2 (c).

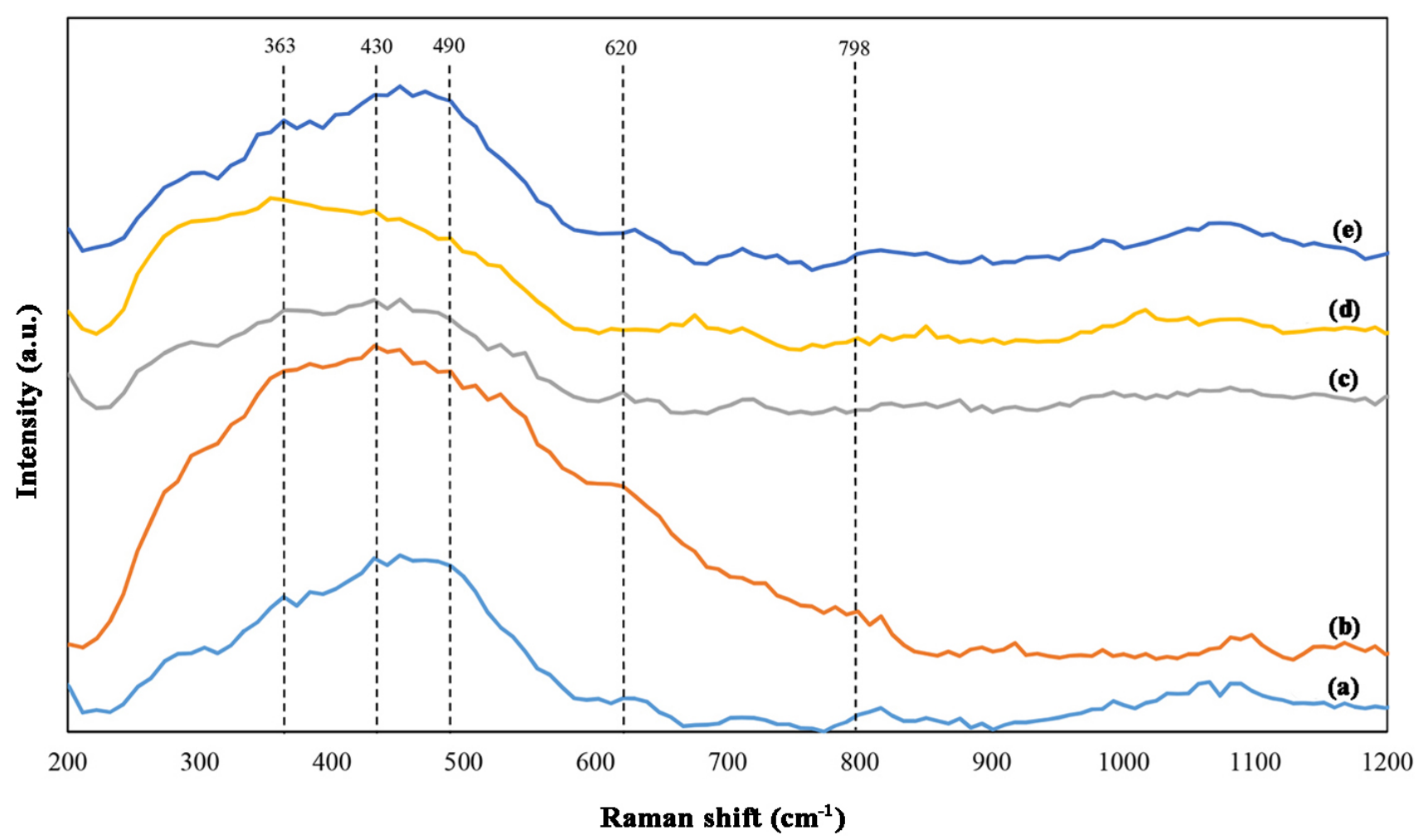

Fig. 2. Raman spectra of the synthesized samples: G0 (a), G0.5 (b), G1 (c), G1.5 (d) and G2 (e).

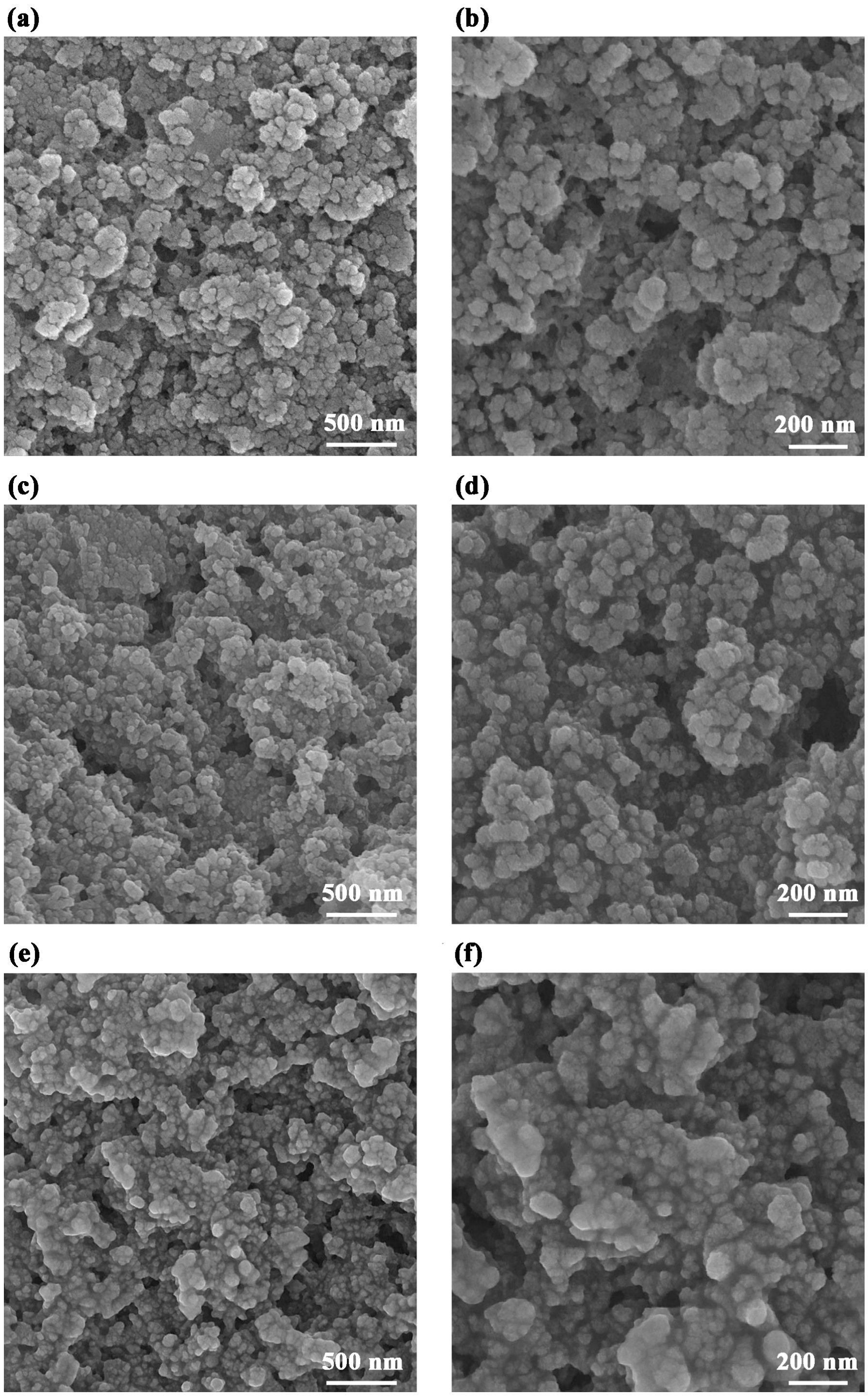

Fig. 3. SEM images of the calcined samples: G0 (a, b), G1 (c, d) and G2 (e, f) in two magnifications.

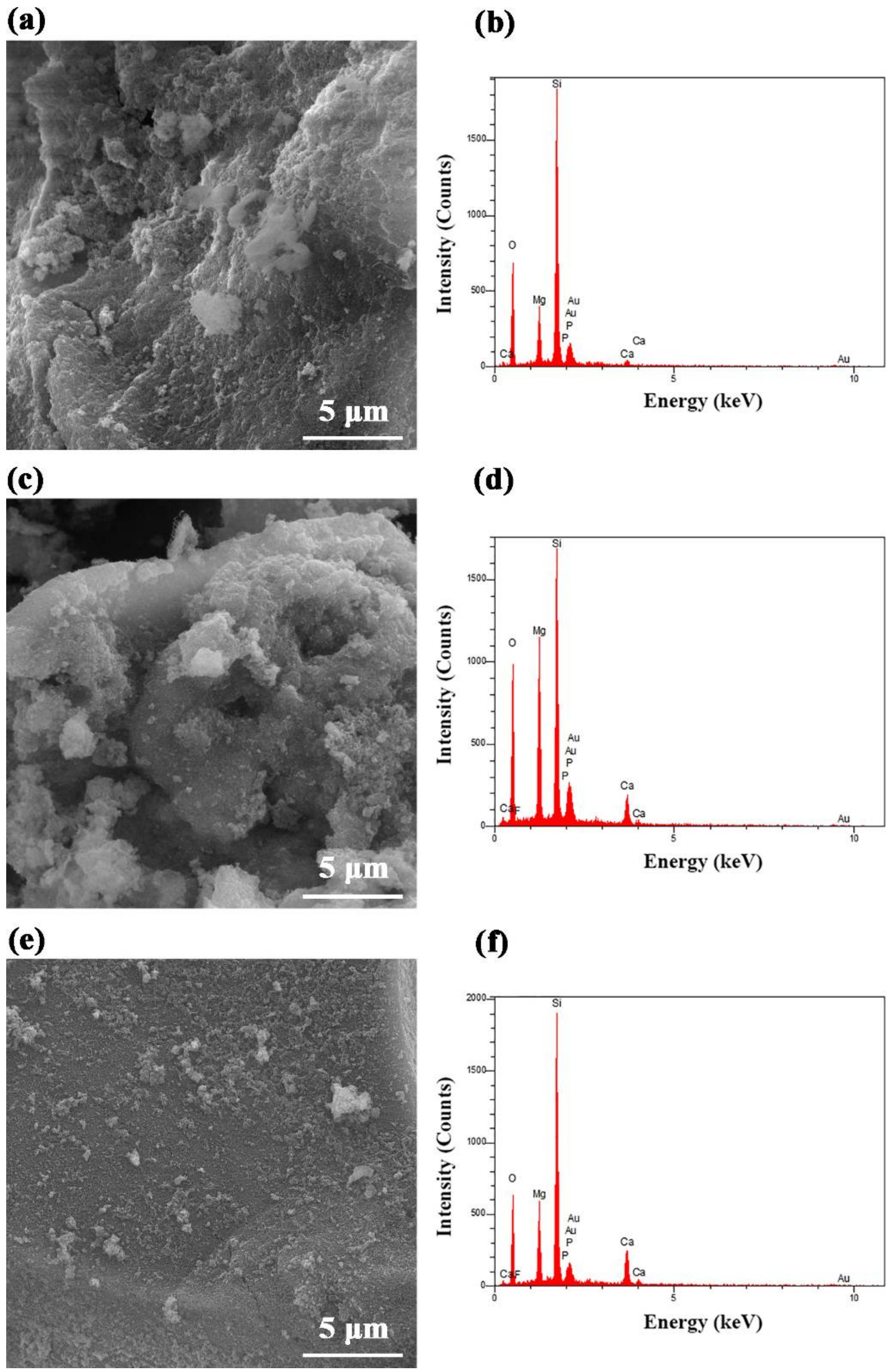

Fig. 4. SEM images and EDS patterns of the samples after immersion in the SBF: G0 (a, b), G1 (c, d) and G2 (e, f).

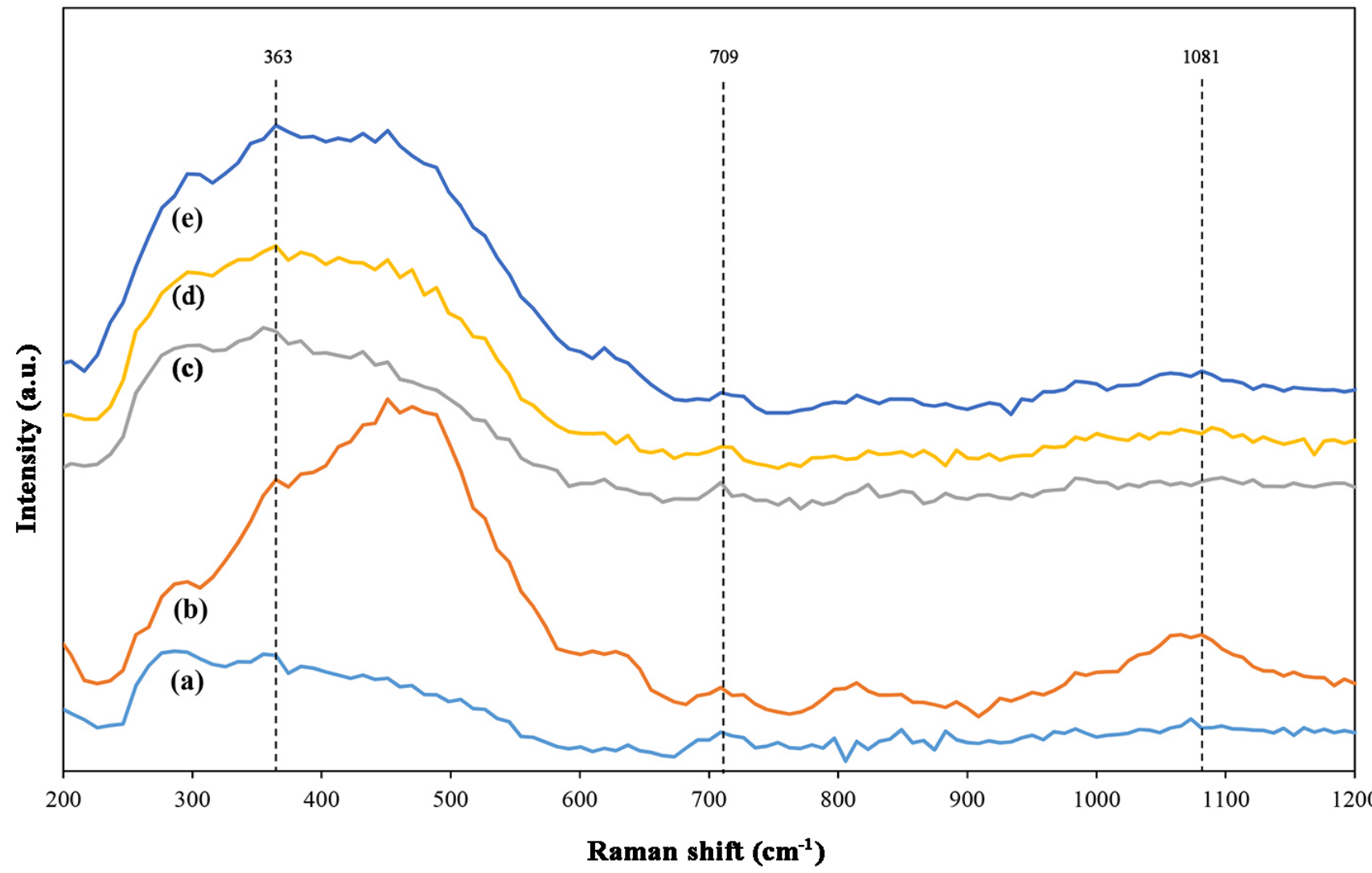

Fig. 5. Raman spectra of the samples after 7 days of immersion in the SBF: G0 (a), G0.5 (b), G1 (c), G1.5 (d) and G2 (e).

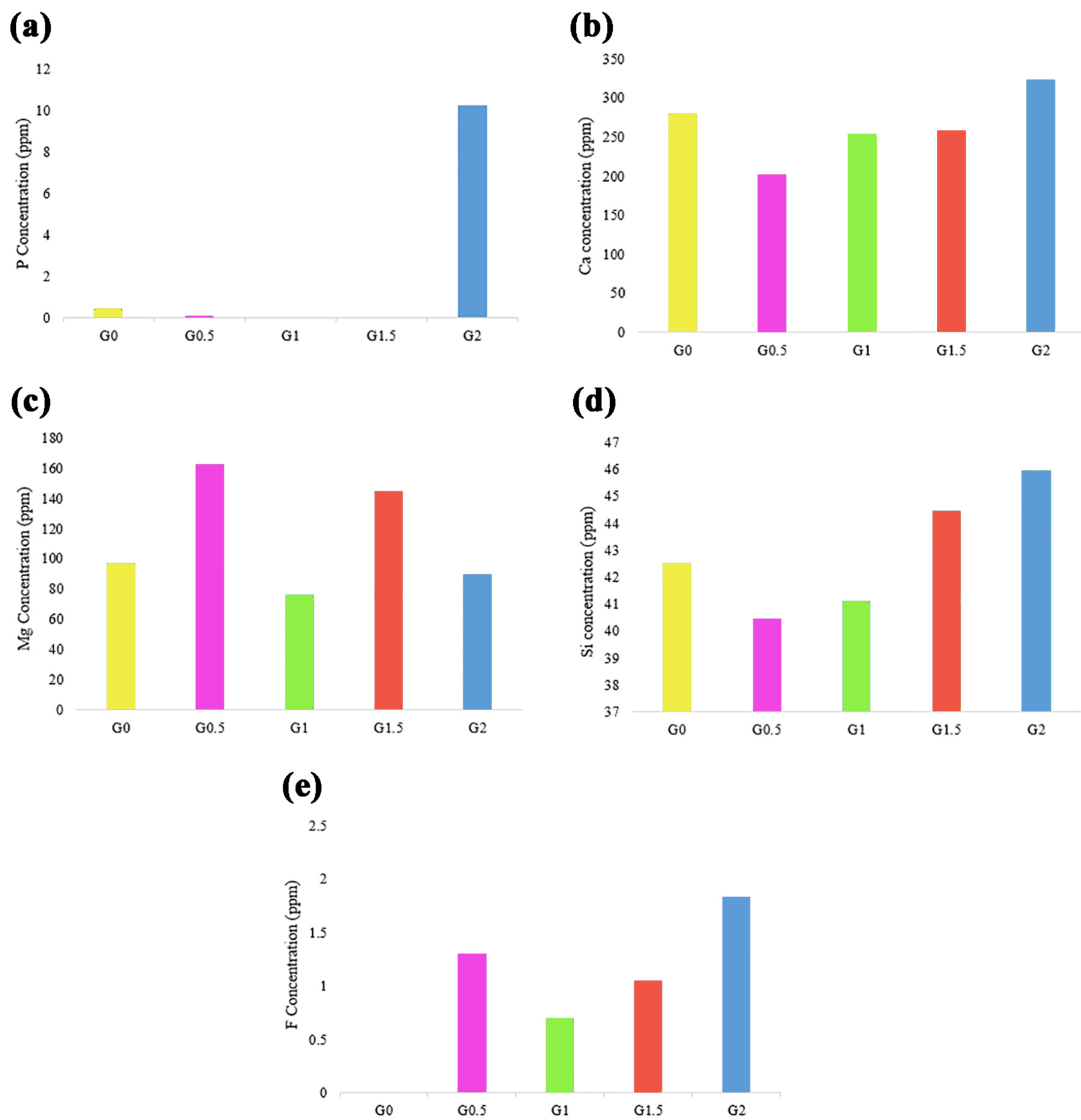

Fig. 6. ICP results of the SBF exposing the synthesized samples: P (a), Ca (b), Mg (c), Si (d) and F (e).

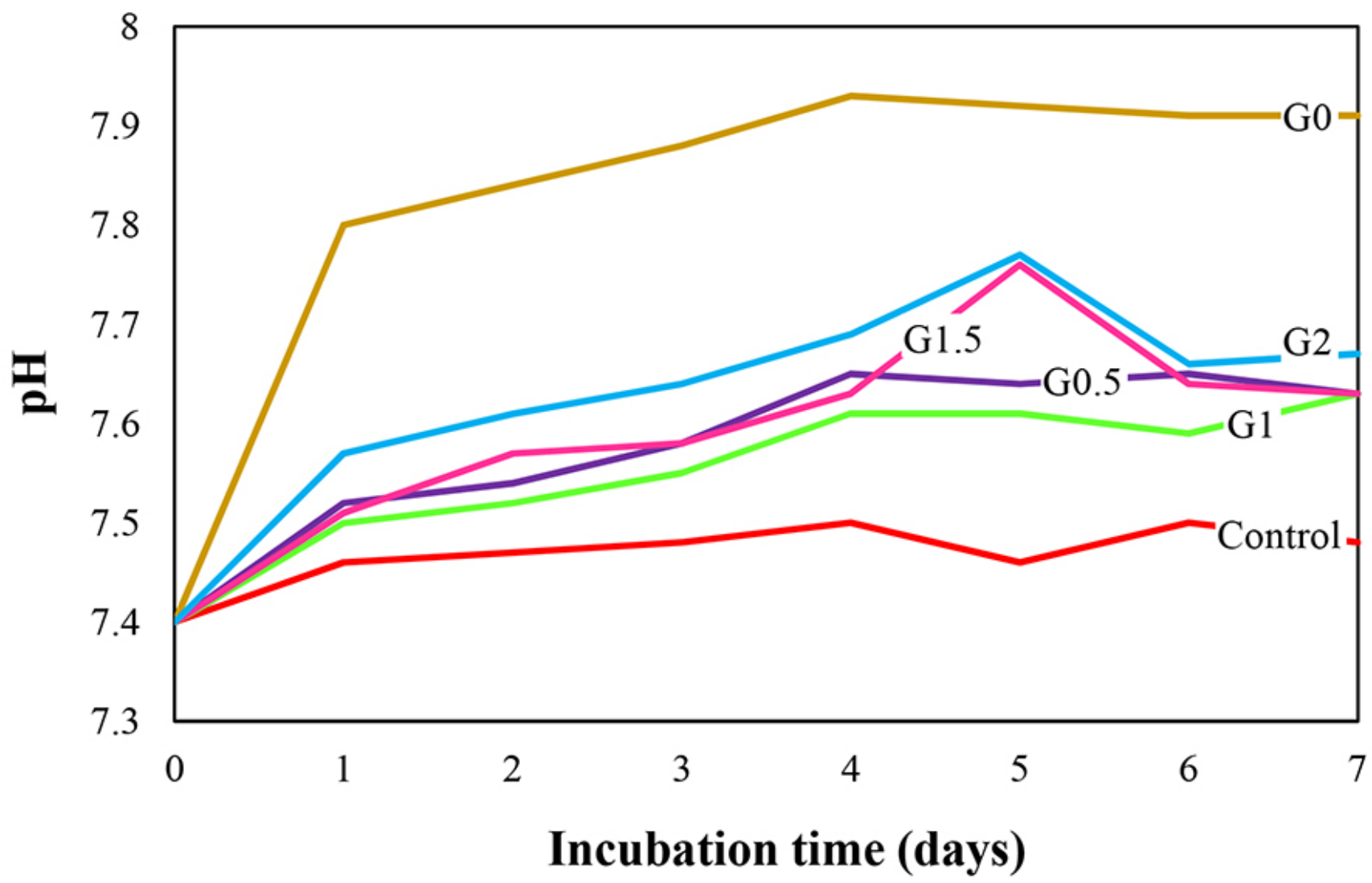

Fig. 7. pH variation of the SBF exposing the synthesized samples.

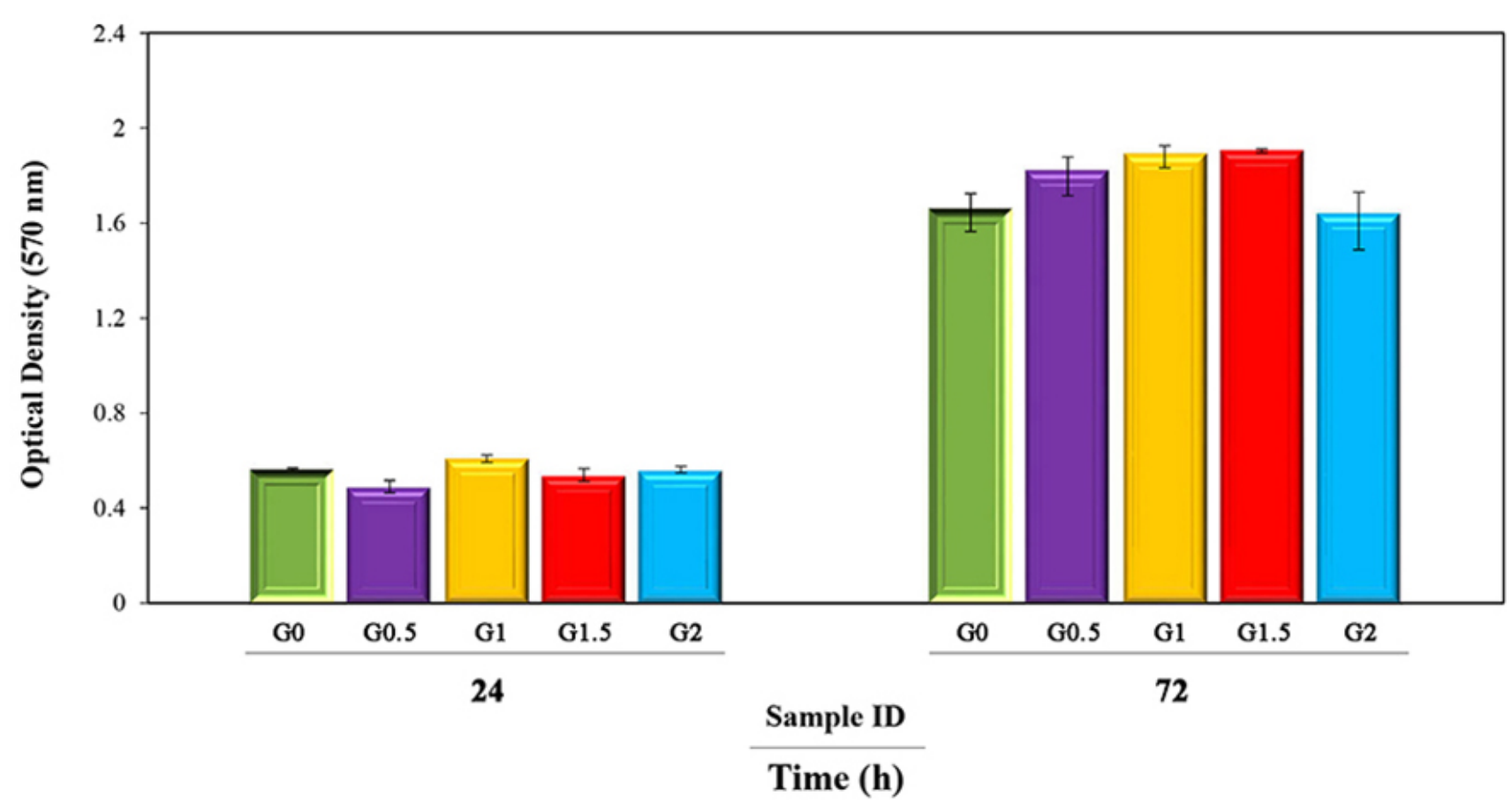

Fig. 8. MMT results of the MG-63 cell cultures on the samples.

Fig. 9. SEM micrograph of the MG-63 cells fixed on the samples: G0 (a), G (b), and G (c).

## Table

Table 1. Amounts of the raw materials used to synthesize the nominated samples (G refers to *glass* in the sample ID column).

| Sample ID | $CaCl_2$ (g) | $MgCl_2$ (g) | $SiCl_4$ (ml) | $MgF_2$ (g) | F (mol%) |
|---|---|---|---|---|---|
| G0 | 2.420 | 2.076 | 5 | 0 | 0 |
| G0.5 | 2.420 | 2.026 | 5 | 0.034 | 0.5 |
| G1 | 2.420 | 1.973 | 5 | 0.068 | 1 |
| G1.5 | 2.420 | 1.921 | 5 | 0.103 | 1.5 |
| G2 | 2.420 | 1.869 | 5 | 0.136 | 2 |